\documentclass[journal,twoside,a4paper]{IEEEtran}

\usepackage{cite}      

\usepackage{amsmath}   
\begin{document}
%
\title{On Feedback and the Classical Capacity of a\\ Noisy Quantum Channel}
%
%
\author{{Garry~Bowen~and~Rajagopal~Nagarajan}
\thanks{G. Bowen is with the Centre for Quantum Computation, Department of Atomic and Laser Physics, Clarendon Laboratory, University of Oxford, Oxford OX1 3PU, UK (e-mail: garry.bowen@merton.ox.ac.uk).}%
\thanks{R. Nagarajan is with the Department of Computer Science, University of Warwick, Coventry CV4 7AL, UK (e-mail: biju@dcs.warwick.ac.uk).}}%
%
%
%
\markboth{Submitted to IEEE Transactions on Information Theory}{Bowen and Nagarajan : On Feedback and the Classical Capacity of a Noisy Quantum Channel}
%



\maketitle

\begin{abstract}
In Shannon information theory the capacity of a memoryless communication channel cannot be increased by the use of feedback from receiver to sender.  In this paper the use of classical feedback is shown to provide no increase in the unassisted classical capacity of a memoryless quantum channel when feedback is used across non-entangled input states, or when the channel is an entanglement--breaking channel.  This gives a generalization of the Shannon theory for certain classes of feedback protocols when transmitting through noisy quantum communication channels.
\end{abstract}


%
\IEEEpeerreviewmaketitle

\section{Introduction}
%
%
%
%

\PARstart{T}{he} theory of quantum information is a generalization of the Shannon theory of information that takes into account the physical nature of the information carrier.  Previously, information was assumed to be encoded in ``classical'' physical states of a system that are distinct and infinitely copyable.  Classical physics remains only an approximation to the underlying quantum nature of matter.  To understand the true limits that the laws of physics will place on our ability to communicate and process information, the quantum behavior of the information carrier must be addressed.

In the classical theory of transmission through noisy channels, the maximum asymptotic rate that information may be transmitted through a discrete memoryless channel (DMC) is given by the Shannon capacity theorem \cite{shannon48a}.  The capacity for a DMC is unchanged even with the inclusion of a number of additional resources, most notably, a noiseless feedback channel \cite{cover}.  This results in a robust measure for the capacity of any memoryless channel.

For noisy quantum channels the situation is somewhat different, and there currently exists a plethora of different capacities for any type of memoryless quantum channel.  By demonstrating relationships between the various capacities, and invariance under the addition of certain resources, the number of different capacities may be reduced to a smaller number.  Any quantum channel may be used to send quantum information, in the form of intact quantum states or entanglement, or used to send classical information encoded in quantum states.  A noisy quantum channel may also be augmented by the use of auxiliary resources, such as one-way or two-way classical side-channels, or prior shared entanglement between sender and receiver.  A channel augmented by a one-way forward classical channel has been shown to have the same quantum capacity as a channel without any classical communication \cite{bennett96,barnum00}.  However, augmentation by classical feedback or a two-way classical channel has been demonstrated to increase the quantum capacity in certain circumstances \cite{bennett96,bowen02a}.  The quantum channel capacities that most resemble the Shannon capacity in both form and behavior appear to be the quantum and classical entanglement--assisted capacities of the channel \cite{adami97,bennett99,bennett01a}.  The formula describing either entanglement assisted capacity is a quantum analogue of the Shannon formula up to a constant factor, and both of these capacities are unchanged by the addition of a noiseless quantum feedback channel \cite{bowen02a}.  For quantum capacities the number of capacities has been reduced to four, $Q$ the unassisted quantum capacity, $Q_{\mathrm{FB}}$ the capacity with a classical feedback channel, $Q_2$ the capacity with a two-way classical side-channel, and $Q_E$ the entanglement--assisted quantum capacity.  The quantum capacities obey the following relationships $Q \leq Q_{\mathrm{FB}} \leq Q_2$ and $Q \leq Q_E$, with known channels for which $Q < Q_{\mathrm{FB}}$ and $Q < Q_E$.  Whether or not there exists a channel for which $Q_{\mathrm{FB}} < Q_2$ remains an open question.

When sending classical information through quantum channels additional resources such as a quantum or classical feedback channel, or prior shared entanglement, may be utilized.  The use of shared entanglement gives rise to the entanglement--assisted capacity $C_E = 2Q_E$, which is equivalent to the classical capacity with a quantum feedback channel.  Whether classical feedback can increase the classical capacity is the question that is addressed in this paper.  It is obvious that the capacity for a channel with feedback is at least as great as the channel without feedback,
\begin{equation}
C_{\mathrm{FB}} \geq C
\end{equation}
and to prove equality it is only necessary to show the reverse inequality.
In this paper the use of classical feedback across non-entangled input states is shown to provide no increase in the classical capacity of noisy quantum channels.  Additionally, if the quantum channel is an entanglement--breaking channel then feedback cannot increase the classical capacity, even for feedback across input states that may be entangled between channel uses.  These results give a partial analogy to the use of feedback in the Shannon theory, and the remaining open case of feedback across entangled input states is examined, with a conjecture that such feedback protocols will not increase the channel capacity.  A proof that feedback cannot increase the classical capacity for the remaining case, of feedback across entangled input states, will result in two well defined classical capacities for memoryless quantum channels, the capacity $C$ where there is no prior shared entanglement, and the entanglement--assisted capacity $C_E$, thus simplifying our current understanding of communication through quantum channels.

\section{The Classical Capacity of a Quantum Channel}

Given a quantum channel $\Lambda$, the Shannon mutual information between a message memory and the average state after transmission is bounded above by the von Neumann mutual information between the states.  A quantum memory for the message states may be constructed, without loss of generality, using known orthogonal pure states $|m^i\rangle \langle m^i|$ corresponding to each message $m^i$.  The memory state acts classically, in the sense that it may measured without error and can be copied arbitrarily many times by measurement and state preparation.  The total state consisting of the message memory and the input states is then,
\begin{equation}
\rho_{MQ} = \sum_i p_i \, |m^i\rangle\langle m^i| \otimes \rho^i_{Q}
\end{equation}
with each state $\rho^i_Q$ corresponding to the message $m^i$, where the message $m^i$ is sent with \textit{apriori} probability $p_i$.  Following transmission of the quantum state through the channel the resultant combined memory and output state is given by,
\begin{equation}
\rho_{MQ'} = \sum_i p_i \, |m^i\rangle\langle m^i| \otimes \Lambda \rho^i_{Q}.
\end{equation}
Any local measurements on the message memory and output state must give a distribution of measurement outcomes such that the Shannon mutual information is bounded by, $I(M:Q') \leq S(M:Q')$ \cite{adami97}, with the von Neumann mutual information between $X$ and $Y$ defined by $S(X:Y) = S(\omega_X) + S(\omega_Y) - S(\omega_{XY})$.  Here, $I(M:Q')$ is the Shannon mutual information between the measurement outcomes on the memory state and the measurement outcomes on the output state of the channel $Q'$.  Expanding the von Neumann mutual information for the alphabet of states $\{ p_i , \rho^i_{Q} \}$, with average density matrix $\rho_Q = \sum_i p_i \rho^i_Q$, obtains the bound \cite{kholevo73},
\begin{align}
S(M:Q') &= S(\rho_M) + S(\Lambda\rho_{Q}) - S\big( ( 1_M \otimes \Lambda_Q ) \rho_{MQ} \big) \nonumber \\
&\leq \sum_i p_i S(\rho^i_M) + S(\Lambda\rho_{Q}) \nonumber \\
&\phantom{=}\: - \sum_i p_i S\big( ( 1_M \otimes \Lambda_Q ) \rho^i_{M} \otimes \rho^i_Q \big) \nonumber \\
&= S(\Lambda \rho_{Q}) - \sum_i p_i S\big( \Lambda \rho^i_Q \big)
\label{eqn:holevo_deriv}
\end{align}
where the inequality follows from the concavity of the conditional entropy $S(\rho_X|\rho_Y) \geq \sum_i p_i S(\rho^i_X|\rho^i_Y)$, with $S(X|Y) = S(\omega_{XY}) - S(\omega_Y)$ the conditional entropy of $X$ given $Y$ \cite{cerf99}.  The last line of (\ref{eqn:holevo_deriv}) follows from the additivity of the entropy for product states $S(\omega_X \otimes \omega_Y ) = S(\omega_X) + S(\omega_Y)$.

The Holevo--Schumacher--Westmoreland (HSW) theorem states that a rate equal to the maximum over all such alphabets is asymptotically attainable with vanishing probability of error \cite{holevo98,schumacher97}.  Therefore the classical information capacity for a quantum channel using a product state alphabet across channel uses, is given by,
\begin{equation}
\chi (\Lambda) = \max_{p_i, \rho^i} \Big[ S(\Lambda\rho) - \sum_i p_i S(\Lambda \rho^i) \Big]
\end{equation}
which can be generalized to give the classical capacity of a quantum channel,
\begin{equation}
C = \lim_{n\rightarrow \infty} \frac{1}{n} \chi(\Lambda^{\otimes n})
\end{equation}
with $\chi(\Lambda^{\otimes n}) = \max_{ \{ p_i, \rho^i \} } \big[ S(\Lambda^{\otimes n} \rho ) - \sum_i p_i S(\Lambda^{\otimes n} \rho^i) \big]$ the HSW capacity for an alphabet in the Hilbert space of maximum dimension $\mathcal{H}^{\otimes n}$, that is, an alphabet that is a product state over uses of blocks of $n$ channels but may be entangled across different channel uses within the same block.  The additivity of the HSW capacity is still an open question, although for certain classes of channels it is known to be additive \cite{king02a,shor02,king03}, $\chi(\Lambda^{\otimes n}) = n\chi(\Lambda)$.

\section{Classical Feedback and Quantum Channels}

To derive the upper bound on the capacity we use the technique of attaching a copy of the message, encoded in mutually orthogonal pure states, to the message states to be transmitted.  In addition, for each use of the feedback channel we add a correlated set of quantum operations, where each operation on the first output is correlated to a trace preserving operation on the second state.  The proof works by induction, showing that for any single step of a feedback protocol the maximum increase of the mutual information between the sender and receiver cannot exceed the HSW bound.  The maximum mutual information generated for a multi-step feedback protocol cannot exceed the sum of the mutual information gained from each step of the protocol, and hence the maximum rate for feedback codes utilized across non-entangled input states or feedback codes for entanglement--breaking channels cannot exceed the classical capacity of the channel without feedback.

\subsection{Feedback Across Product Input States}

In order to prove the result for the most general type of protocol for product state inputs, the channels may be assumed to be of the form $\Omega \otimes \Lambda$, where for multiple use of a single channel $\Phi$ we can assume $\Omega = \Phi^{\otimes n}$ and $\Lambda = \Phi^{\otimes m}$, with $m$ and $n$ arbitrary.

The initial total state is of the form,
\begin{equation}
\rho_{MQ_1Q_2} = \sum_i p_i \, |m^i\rangle\langle m^i| \otimes \rho^i_{Q_1} \otimes \rho^i_{Q_2}
\end{equation}
with message $m^i$ being sent with \textit{apriori} probability $p_i$.  The state $Q_1$ is then sent through the first channel to produce the new state,
\begin{equation}
\rho_{MQ'_1Q_2} = \sum_i p_i |m^i\rangle\langle m^i| \otimes \Omega\rho^i_{Q_1} \otimes \rho^i_{Q_2}.
\label{eqn:first_use_state}
\end{equation}
This is then followed by the \textit{feedback operation}, which without loss of generality may be represented by classically correlated operations on the feedback state $\mathcal{B}_{Q_1}$ and a combined operation on the memory state and next state to be transmitted $\mathcal{A}_{MQ_2}$, where the copy of the message memory must remain invariant under the operation $\mathcal{A}_M$.  The total state after the feedback operation and transmission of the second state through the channel is then,
\begin{equation}
\rho_{MQ'_1Q'_2} = \sum_{i} p_i |m^i\rangle\langle m^i| \otimes \big( {1}_{Q_1} \otimes \Lambda_{Q_2}\big) \omega^{\, i}_{Q'_1Q_2}
\label{eqn:final_state}
\end{equation}
where,
\begin{align}
\omega^{\, i}_{Q'_1Q_2} &= \mathrm{Tr}_M \bigg[ \sum_{jk} A^{(j)k}_{MQ_2}\big(|m^i\rangle \langle m^i| \otimes \rho^i_{Q_2}\big) A^{(j)k\dag}_{MQ_2} \nonumber \\
&\phantom{=}\qquad \qquad \otimes B^{j}_{Q_1}\Omega\rho^i_{Q_1}(B^{j}_{Q_1})^{\dag} \bigg] \nonumber \\
&= \sum_{jk} B^{j}_{Q_1}\Omega\rho^i_{Q_1}B^{j\dag}_{Q_1} \otimes A^{(ij)k}_{Q_2} \rho^i_{Q_2} A^{(ij)k\dag}_{Q_2}.
\end{align}
For each $i$ the state is an action utilizing only local operations and classical communication (LOCC).  Any LOCC action on a separable state necessarily results in a separable state, hence each $\omega^i_{Q'_1Q_2}$ is separable, and the convex sum of these separable states $\omega_{Q'_1Q_2}$ is also separable.  As each state following the feedback operation is separable, it may be written as a convex sum over product states,
\begin{equation}
\omega^i_{Q'_1Q_2} = \sum_j q_j \, \omega^{ij}_{Q'_1} \otimes \omega^{ij}_{Q_2}
\end{equation}
where $q_j \geq 0$ and $\sum_j q_j = 1$.
The total state in (\ref{eqn:final_state}), following both the feedback operation and the second transmission, may then be rewritten in the form,
\begin{equation}
\rho_{MQ'_1Q'_2} = \sum_{ij} p_i q_j \, |m^i\rangle\langle m^i| \otimes \omega^{ij}_{Q'_1} \otimes \Lambda \omega^{ij}_{Q_2}.
\label{eqn:total_separable}
\end{equation}

The mutual information between the message memory and the combined output states may be rephrased in terms of the reduced mutual information and the conditional mutual information,
\begin{equation}
S(M:Q'_{1}Q'_{2}) = S(M:Q'_1) + S(M:Q'_2|Q'_1)
\label{eqn:mutual_exp}
\end{equation}
with the conditional mutual information defined in terms of the conditional entropies by,
\begin{equation}
S(M:Q'_2|Q'_1) = S(Q'_2|Q'_1) - S(Q'_2|MQ'_1).
\label{eqn:cond_mut_def}
\end{equation}
The conditional information for quantum states differs from the classical counterpart, in that, the two terms on the right hand side of (\ref{eqn:cond_mut_def}) can each be negative, but only for entangled states.
The first term on the right hand side of (\ref{eqn:mutual_exp}) may be explicitly written as,
\begin{align}
S(M:Q'_1) &= S(\rho_M) + S(\mathcal{B}\Omega \rho_{Q_1}) - S(\mathcal{B}\Omega \rho_{MQ_{1}}) \nonumber \\
&\leq S(\rho_M) + S(\Omega \rho_{Q_1}) - S(\Omega \rho_{MQ_{1}}) \nonumber \\
&\leq \chi_{Q_1}
\label{eqn:first_bound}
\end{align}
where $\chi_{Q_1}$ is the HSW capacity of the first channel.  The first inequality follows from the fact that any quantum operation $\mathcal{B}$ acting on part of a bipartite state cannot increase the von Neumann mutual information \cite{lindblad75}, and the second inequality follows from the definition of $\chi_{Q}$.  The first inequality in (\ref{eqn:first_bound}) incorporates any information gained from the measurement outcome during the feedback protocol.  This may be seen by attaching an initially pure ancilla state $|0_A\rangle\langle 0_A|$ which after the new operation on the output state and ancilla $\mathcal{B}_{Q'_1A}$, gives a classical copy of the measurement outcome in the ancilla state $B^j \rho_{Q'_1}B^{j\dag} \otimes |j_A\rangle \langle j_A|$.  The mutual information between the message memory and the output state combined with the ancilla following the measurement operation $\mathcal{B}_{Q'_1A}$, must also necessarily be less than the initial mutual information between the message memory and the initial output state and ancilla state.  Because the ancilla is initially in a product state, the first inequality in (\ref{eqn:first_bound}) then follows from the additivity of the entropies of product states $S\big(\rho_{Q'_1}\otimes |0_A\rangle\langle 0_A|\big) = S(\rho_{Q'_1})$.

Obtaining the required bound on the final term of (\ref{eqn:mutual_exp}) is only slightly more difficult, and we begin by expanding the terms according to the basic definition in terms of the conditional von Neumann entropies, such that,
\begin{align}
S(M:Q'_2|Q'_1) &= S(\rho_{Q'_2}|\rho_{Q'_1}) - S(\rho_{Q'_2}|\rho_{MQ'_1}) \nonumber \\
&\leq S(\rho_{Q'_2}) - S(\rho_{Q'_2}|\rho_{MQ'_1})
\label{eqn:conditional_mutual}
\end{align}
where the inequality follows from the fact that conditioning cannot increase the entropy in the first term \cite{cerf99}.  As the conditional entropy is concave, the second term $-S(\rho_{Q'_2}|\rho_{MQ'_1})$ is convex, and from the decomposition of each of the states $\omega^i_{Q'_1Q_2}$ as a separable state, the bound
\begin{align}
-S(\rho_{Q'_2}|\rho_{MQ'_1}) &= S\Big(\sum_{ij} p_i q_j \rho^i_M \otimes \omega^{ij}_{Q'_1}\Big) \nonumber \\
&\phantom{=}\:- S\Big(\sum_{ij} p_i q_j \rho^i_M \otimes \omega^{ij}_{Q'_1} \otimes \Lambda\omega^{ij}_{Q_2} \Big) \nonumber \\
&\leq \sum_{ij} p_i q_j S\Big(\rho^i_M \otimes \omega^{ij}_{Q'_1}\Big) \nonumber \\
&\phantom{=}\:- \sum_{ij} p_i q_j S\Big(\rho^i_M \otimes \omega^{ij}_{Q'_1} \otimes \Lambda\omega^{ij}_{Q_2} \Big) \nonumber \\
&= -\sum_{ij} p_i q_j S\big( \Lambda\omega^{ij}_{Q_2} \big)
\label{eqn:cond_term2}
\end{align}
is obtained.
Substituting (\ref{eqn:cond_term2}) into (\ref{eqn:conditional_mutual}) then gives,
\begin{align}
S(M:Q'_2|Q'_1) &\leq S\big(\Lambda\omega_{Q_2}\big) - \sum_{ij} p_i q_j S\big( \Lambda\omega^{ij}_{Q_2} \big) \nonumber \\
&\leq \chi_{Q_2}
\label{eqn:second_chi}
\end{align}
with $\rho_{Q'_2} = \Lambda \omega_{Q_2}$, and the decomposition,
\begin{equation}
\sum_{ij} p_i q_j \, \omega^{ij}_{Q_2} = \omega_{Q_2}.
\end{equation}
The second inequality in (\ref{eqn:second_chi}) follows from the fact that the HSW capacity is the maximum over all such ensembles.
Hence, the total capacity of the feedback protocol across the channels is bounded above by the separate HSW capacities of the channels,
\begin{equation}
\chi^{\mathrm{FB}}_{Q_1Q_2} \leq S(M:Q'_1Q'_2) \leq \chi_{Q_1} + \chi_{Q_2}.
\label{eqn:feedback_bound}
\end{equation}
The same result follows for an arbitrary feedback protocol, across any product states inputs, by induction.

\subsection{Feedback Across Non--entangled Input States}

The upper bound on any feedback protocol in the previous section is easily extended to include the class of input states that are non-entangled, that is, all separable states.  This is due to the fact that any separable input states will remain separable after any LOCC feedback operation, and can therefore be written in the form of (\ref{eqn:total_separable}).  Therefore feedback across non-entangled input states cannot increase the maximum asymptotic rate at which classical information may be sent through a memoryless quantum channel.

\subsection{Entanglement--breaking Channels}

An entanglement--breaking quantum channel is one which cannot transmit entanglement.  If part of any bipartite entangled state is transmitted through the channel, then the bipartite state following transmission is necessarily separable.  Explicitly, for any initial state $|\phi_{RQ}\rangle$, the output $\rho_{RQ'}$ given by,
\begin{equation}
\rho_{RQ'} = \big( 1_R \otimes \Lambda_Q \big) |\phi_{RQ}\rangle \langle \phi_{RQ}|
\end{equation}
is always separable.  The classical capacity for entanglement--breaking channels has previously been shown to be additive, and hence the classical information capacity for such channels is simply $C = \chi (\Lambda)$ \cite{shor02}.

The proof that feedback cannot increase the classical capacity of an entanglement--breaking channel is straightforward.  After the first state $Q_1$ is sent through an entanglement--breaking channel, the total state is necessarily separable and may be written in the form,
\begin{equation}
\rho_{MQ'_1Q_2} = \sum_k \pi_k \, \rho^k_{MQ'_1} \otimes \rho^k_{Q_2}
\end{equation}
for which the feedback operation will result in a new separable state,
\begin{equation}
\omega_{MQ'_1Q_2} = \sum_k \eta_k \, \omega^k_{MQ'_1} \otimes \omega^k_{Q_2}.
\label{eqn:first_use_state2}
\end{equation}
Following transmission of the second state, the mutual information between $MQ'_1$ and $Q'_2$ is bound by,
\begin{align}
S(MQ'_1:Q'_2) &= S(M) + S(Q'_2) - S(MQ'_1Q'_2) \nonumber \\
&\leq S(Q'_2) + \sum_k \eta_k S(\omega^k_{MQ'_1}) \nonumber \\
&\phantom{=}\:- \sum_k \eta_k S(\omega^k_{MQ'_1} \otimes \Lambda\omega^k_{Q_2}) \nonumber \\
&= S(\Lambda\omega_{Q_2}) - \sum_k \eta_k S(\Lambda\omega^k_{Q_2}) \nonumber \\
&\leq \chi_{Q_2}
\end{align}
with $\omega_{Q_2} = \sum_k \eta_k \, \omega^k_{Q_2}$.  The total mutual information over the two channels is therefore bound by,
\begin{align}
S(M:Q'_1Q'_2) &= S(M:Q'_1) + S(M:Q'_2|Q'_1) \nonumber \\
&\leq S(M:Q'_1) + S(MQ'_1:Q'_2) \nonumber \\
&\leq \chi_{Q_1} + \chi_{Q_2}
\label{eqn:feed_subadd2}
\end{align}
and consequently feedback cannot increase the classical information capacity of entanglement--breaking channels.

The derivation from (\ref{eqn:first_use_state2}) to (\ref{eqn:feed_subadd2}) is essentially a less detailed, but otherwise almost identical, version of (\ref{eqn:first_use_state}) to (\ref{eqn:feedback_bound}), with the major difference being that in (\ref{eqn:first_use_state}) the alphabet states are product states rather than separable states after the use of the channel.
It may be noted, however, that for a feedback protocol to possibly exceed the bound for the non-feedback capacity the \textit{average total states}, $\rho_{MQ'_1Q_2}$ and $\omega_{MQ'_1Q_2}$, must remain entangled between the states that have been sent and the states held by the sender, at some step of the protocol.  Any channel for which all possible ensembles do not obey this property must therefore have $C_{\mathrm{FB}} = C$, which is the defining characteristic of any entanglement--breaking channel.

\section{Conclusion}

The use of classical feedback for the transmission of classical information through a memoryless quantum channel has been shown to give no increase in the capacity of the channel when the feedback is used across non-entangled input states.  Additionally, it has been shown that feedback cannot increase the classical capacity of entanglement--breaking channels.  The question of whether or not feedback can increase the capacity of memoryless quantum channels when used across entangled input states remains open.

\end{document}